# Positron Processes in the Sun

**Nat Gopalswamy***

Solar Physics Laboratory, NASA Goddard Space Flight Center, Greenbelt, MD 20771, USA
* Correspondence: nat.gopalswamy@nasa.gov

**Abstract:** Positrons play a major role in the emission of solar gamma-rays at energies from a few hundred keV to >1 GeV. Although the processes leading to positron production in the solar atmosphere are well known, the origin of the underlying energetic particles that interact with the ambient particles is poorly understood. With the aim of understanding the full gamma-ray spectrum of the Sun, I review the key emission mechanisms that contribute to the observed gamma-ray spectrum, focusing on the ones involving positrons. In particular, I review the processes involved in the 0.511 MeV positron annihilation line and the positronium continuum emissions at low energies, and the pion continuum emission at high energies in solar eruptions. It is thought that particles accelerated at the flare reconnection and at the shock driven by coronal mass ejections are responsible for the observed gamma-ray features. Based on some recent developments I suggest that energetic particles from both mechanisms may contribute to the observed gamma-ray spectrum in the impulsive phase, while the shock mechanism is responsible for the extended phase.

## 1. Introduction

Positrons, the antiparticle of electrons, were proposed theoretically by Dirac and were first detected by Anderson in 1933 [1]. Positrons are extensively used in the laboratory for a myriad of purposes (see review by Mills [2]). Astrophysical processes involving positrons have been found in the interstellar medium [3], and galactic bulge and disk [4]. Positrons are commonly found in the Sun [5]. The proton–proton chain, which accounts for most of the energy release inside the Sun involves the emission of a positron when two protons collide to form a deuterium nucleus. There are plenty of electrons present in the solar core, so the positrons are immediately annihilated with electrons and produce two gamma-ray photons. In one proton–proton chain, two positrons are emitted and hence contribute a total of four photons in addition to the two photons emitted during the formation of a helium-3 ($^3$He) nucleus from the fusion of deuterium nucleus with a proton.

High-energy particles exist in the solar atmosphere, energized during solar eruptions. Solar eruptions involve flares and coronal mass ejections (CMEs). A process known as magnetic reconnection taking place in the solar corona is thought to be the process which converts energy stored in the stressed solar magnetic fields into solar eruptions (see Figure 1 for a schematic of a typical eruption). One part of the released energy heats the plasma in the eruption region, while another goes to energize electrons and ions. Electromagnetic radiation from radio waves to gamma-rays produced by the energized electrons and protons by various processes is known as a flare. Acceleration in the reconnection region is referred to as flare acceleration or stochastic acceleration. CMEs also carry the released energy as the kinetic energy of the expelled magnetized plasma with a mass as high as $10^{16}$ g and speeds exceeding 3000 km/s. Such fast CMEs drive fast-mode magnetohydrodynamic shocks that can also energize ambient electrons and ions to very high energies. Diffusive shock acceleration and shock drift acceleration are the mechanisms by which particles are energized at the shock (see [7] for a review on acceleration mechanisms during solar eruptions). The flare and shock accelerations were referred to as first- and second-phase accelerations during an eruption by Wild et al. [8]. Accelerated particles have access to open and closed magnetic structures associated with the eruption resulting in a number of electromagnetic emissions via different emission mechanisms. Energetic particles escaping along interplanetary magnetic field lines are detected as solar energetic particle (SEP) events by particle detectors in space and on ground. These particles, originally known as solar cosmic rays, were first detected by Forbush [9] in the 1940s.

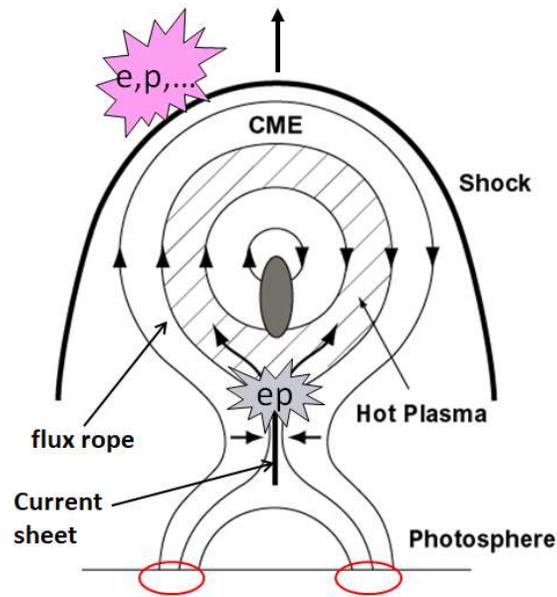

**Figure 1.** Schematic of a solar eruption and the sites of particle acceleration (e,p, …): one in the current sheet formed low in the corona and the other on the surface of the shock driven by the coronal mass ejection (CME) flux rope. The arrows toward the current sheet indicate the reconnection inflow while the ones diverging away indicate the outflow. The red ellipses in the photosphere represent the feet of flare loops where accelerated particles precipitate and produce flare radiation. Particles from the shock propagate away from the Sun and are detected as energetic particle events in space. The dark ellipse inside the flux rope represents a prominence that erupted along with the CME (adapted from Gopalswamy [6]).

Invoking the copious production of energetic particles from the Sun, Morrison [10] suggested that the active Sun must be a source of gamma-rays. He listed electron-positron annihilation as one of the processes expected to produce gamma-ray emission from cosmic sources [10]. Elliot [11] suggested that "positive electrons" from muon decay should lead to detectable 0.5 MeV gamma-ray line emission. Lingenfelter and Ramaty [12] performed detailed calculations of gamma-ray emission processes from the Sun. Chupp et al. [13] identified for the first time the positron annihilation radiation at 0.5 MeV along with other nuclear lines during the intense solar flares of 1972 August 4 and 7 using data from the Gamma-ray Monitor onboard NASA's Seventh Orbiting Solar Observatory (OSO-7) mission.

**2. Mechanisms of Positron Production**

Positrons are predominantly produced by three processes: (i) emission from radioactive nuclei, (ii) pair production by nuclear deexcitation, and (iii) decay of positively charged pions ($\pi^+$) that take place when ions accelerated in the corona interact with the ions in the photosphere/chromosphere. Kozlovsky et al. [14] list 156 positron-emitting radioactive nuclei resulting from the interaction of protons and $\alpha$-particles (helium nuclei) with 12 different elements and their isotopes. The most important positron-emitting radioactive nuclei that result from the interaction of protons (p) and $\alpha$- particles with carbon ($^{12}$C, $^{13}$C), nitrogen ($^{14}$N, $^{15}$N), and oxygen ($^{16}$O, $^{18}$O) are listed in Table 1. In the interaction of p and $\alpha$ with $^{16}$O, the oxygen nucleus, ends up in the excited state ($^{16}$O*) of 6.052 MeV; this nucleus deexcites by emitting an electron-positron pair with a lifetime of 0.096 ns. Another such excited nucleus is $^{40}$Ca* with a lifetime of 3.1 ns. Kozlovsky et al. [15] list another set of 23 positron emitters produced when accelerated $^{3}$He interact with targets such as $^{12}$C, $^{14}$N, $^{16}$O, $^{20}$Ne, $^{24}$Mg, $^{28}$Si, and $^{56}$Fe. The radioactive nuclei have a lifetime ranging from a tenth of a nanosecond to ~1 million years (see [14] for a list). There are 26 radioactive nuclei with a lifetime ≲10s.

Table 1. Important interactions and the resulting radioactive nuclei.

| Interaction | Target nuclei | Radioactive nuclei |
|---|---|---|
| p - carbon | 12C | 11C, 12N, 10C, 13N |
| p - carbon | 13C | 13N |
| $\alpha$ - carbon | 12C | 11C, 15O |
| p - nitrogen | 14N | 11C, 13N, 14O |
| p - nitrogen | 15N | 15O |
| $\alpha$ - nitrogen | 14N | 17F, 13N, 11C |
| p - oxygen | 16O | 11C, 13N, 15O, 16O* |
| p - oxygen | 18O | 18F |
| $\alpha$ - oxygen | 16O | 19Ne, 18F, 15O, 13N, 11C, 16O* |

Note: *Nucleus in excited state.

Positrons from radioactive nuclei have an energy of several hundred keV, while those from $\pi^+$ decay have much higher energy (up to hundreds of MeV). Almost all the positrons emitted by radioactive nuclei and a major fraction of those produced by $\pi^+$ decay (~80%) slow down to thermal levels (tens of eV) before directly annihilating or forming a positronium (Ps) atom by capturing an electron. The formation of Ps in this way is via radiative recombination. Ps can also be formed due to charge exchange with H and He atoms. Positronium is a hydrogen-like atom (with the proton replaced by a positron). There are two types of Ps, known as orthopositronium (O-Ps) and parapositromium (P-Ps), depending on how the spins of the positron and electron are oriented. In O-Ps, the electron and positron spins are in the same direction (triplet state); in P-Ps, the spins are oppositely directed (singlet state). O-Ps and P-Ps decay into 3 and 2 photons, respectively. O-Ps is formed preferentially: 75% of the time compared to 25% of the time for P-Ps [16]. Four key processes that determine the fate of the positrons produced in the solar atmosphere are discussed in [5]. These processes involve the interaction of positrons with the ambient hydrogen and helium in redistributing their energy, formation and quenching of Ps, and the ultimate production of gamma-rays by direct annihilation or via Ps. Direct annihilation of positrons can occur with free and bound (in H and He) electrons in the ambient medium. Positronium quenching occurs resulting in the emission of second-generation positrons when Ps collides with electrons, H, H$^+$, and He$^+$. Another quenching process is the conversion of O-Ps to P-Ps when O-Ps collides with electrons and H.

Pions ($\pi^0$ and $\pi^\pm$) are created when accelerated protons and $\alpha$-particles from the corona collide with those in the chromosphere/photosphere. A detailed list of possible interactions (p-p and p-$\alpha$) are listed in Murphy et al. [17]. High-energy positrons are primarily emitted from the decay of $\pi^+$ into positive muons ($\mu^+$), which decay into positrons. In a similar reaction, negative pions ($\pi^-$) decay into negative muons ($\mu^-$), which decay into electrons. $\pi^0$ decays into 2 gamma-rays most of the time (98.8%). In the remaining 1.2% of cases, $\pi^0$ decays into an electron positron pair and a gamma-ray. The rest energy of neutral pions is 135 MeV, while that of charged pions is 139.6 MeV. To produce these particles, the accelerated protons need to have high energies, exceeding ~300 MeV. The pions are very short-lived ($\pi^0$: $10^{-16}$ s; $\pi^\pm$: $2.6\times10^{-8}$ s), while the muons live for a couple of microseconds ($2.2\times10^{-6}$ s). The >300 MeV protons needed for pion production seem to be accelerated both in the flare reconnection and CME-driven shocks (see Figure 1).

## 3. Gamma-Rays Due to Positrons

Nonthermal electromagnetic emission produced by charged particles is one of the key evidences for particle acceleration in the Sun (see, e.g., [18] for a review). Energetic electrons are readily inferred from nonthermal emission they produce at wavelengths ranging from millimeters to kilometers. Theories of radio emission have helped us understand the acceleration mechanism for the electrons and the radio emission mechanism. Energetic electrons also produce hard X-rays and gamma-rays. On the other hand, the electromagnetic indicators of energetic ions are limited to gamma-rays, ranging from a few hundred keV to >1 GeV. Positrons, which are produced by various processes noted in section 2, contribute to the gamma-ray emission from the Sun at various energies via different processes.

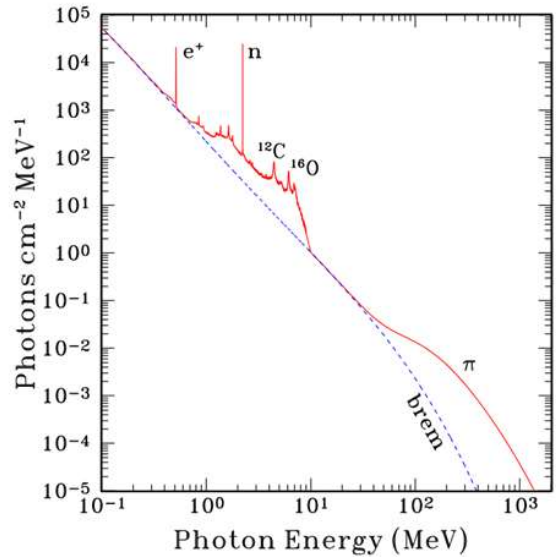

**Figure 2.** Overall theoretical spectrum of gamma-ray emission from the Sun from 0.1 MeV to 2 GeV. The blue dashed line labeled "brem" represents the contribution from the bremsstrahlung of energetic electrons accelerated during solar eruptions. The red line represents computed spectrum taking into account of all possible processes that contribute to gamma-ray emission. At energies below ~10 MeV there is a quasi-continuum with several lines superposed. e+ and n represent the 0.511 MeV positron annihilation line and the 2.223 MeV neutron capture line. $^{12}$C and $^{16}$O mark the next intense lines produced by nuclear deexcitation. The pion continuum at high energies is denoted by π, which involves contribution from both neutral and charged pions. The energy for these emission components is from energetic particles (electrons and ions) accelerated in the solar corona during solar eruptions (adapted from Ramaty and Mandzhavidze [19]).

Figure 2 shows the total gamma-ray spectrum from the Sun, exhibiting both line emission and continuum as predicted [10]. This spectrum is constructed from all possible processes that emit gamma-rays during solar eruptions [19]. At low energies (< 10 MeV), there are many lines superposed on the electron bremsstrahlung continuum, the lowest being the 0.511 MeV line (positron annihilation, marked e+). The next narrow line is the neutron capture line at 2.223 MeV. When energetic protons spallate ambient nuclei, neutrons are produced and emitted over a broad angular distribution; the downward neutrons slow down and are captured by a proton in the ambient medium forming a deuterium nucleus and releasing the binding energy as the 2.223 MeV line. The other lines are due to nuclear deexcitation of varying widths due to various combinations of incumbent and insurgent ions. The continuum emission has several components. Below the 0.511 MeV line, there is a weak Ps continuum that merges with the strong primary electron bremsstrahlung continuum. Primary electrons are those arriving from the acceleration site in the corona, as opposed to secondary electrons, which are produced in the chromosphere/photosphere due to the impact of accelerated ions from the corona (e.g., via π- decay). The quasi-continuum between 0.7 MeV and 10 MeV (on which the discrete lines are superposed) is due to a multitude of broad nuclear lines caused by insurgent heavy ions interacting with ambient H and He nuclei. If the primary electron bremsstrahlung continuum is hard, it can be detected above background even at energies exceeding 10 MeV. When there is significant pion production both a broad line like feature centered near 70 MeV from neutral pions and a hard, secondary positron bremsstrahlung continuum can also be detected above 10 MeV (see below).

*3.1. The 0.511 MeV Gamma-Ray Line*

The width of the solar annihilation line can range from ~1 keV to 10 keV full width at half maximum (FWHM) depending on the ambient conditions of the medium in which the annihilation takes place: e.g., temperature, density, and the ionization state [20]. Detailed calculations and comparison with observations have confirmed that the 0.511 MeV line is produced over a range of these parameters in the chromosphere/photosphere region [5]. One of the major contributors to the line width is the temperature in the region of annihilation because the ambient electrons have a distribution of speeds at a given temperature. The environmental conditions also determine the formation and destruction of Ps. Parapositronium annihilates emitting two 0.511 MeV photons (2γ decay), with lifetime ~0.125 ns. On the other hand, orthopositronium annihilates in three 341 keV photons ((3γ decay) with a lifetime ~142 ns [21]. Depending on the initial energy of the positron capturing an electron, the 3γ decay results in a gamma-ray continuum at energies below the annihilation line (see Figure 2). The flux ratio of the 3γ continuum (from O-Ps) to the 2γ line (from direct annihilation with free and bound electrons, and the decay of P-Ps) is an important parameter that can be used to infer the properties of the ambient medium (density, temperature, and ionization state). Murphy et al. [5] found that the flux ratio at a given temperature in an ionized medium remains constant up to an ambient hydrogen density of ~$10^{13}$ cm$^{-3}$ and then rolls over to values lower by 2–3 orders of magnitude at densities ~$10^{17}$ cm$^{-3}$. The constant value depends on the ambient temperature starting from ~3 for 2000 K and decreasing to ~0.001 at 10 MK. For a neutral atmosphere, the temperature dependence is weak: the constant value of the flux ratio is ~5 at densities below ~$10^{14}$ cm$^{-3}$ and rolling over to ~0.008 at an ambient density of ~$10^{17}$ cm$^{-3}$.

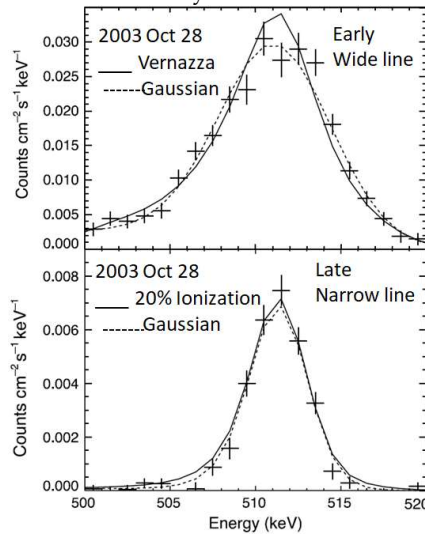

**Figure 3.** Profiles of the 0.511 MeV annihilation line during the 2003 October 28 event observed by the Reuven Ramaty High Energy Solar Spectroscopic Imager (RHESSI). (Top) During the early phase (first 2 minutes of observation) the data points are fitted with a Vernazza atmosphere [22] at a temperature of 6000 K and a Gaussian with a width of ~6.7 keV corresponding to a temperature of (3–4) ×$10^5$ K . (Bottom) During the late phase (last 16 minutes of observation), the data are fitted with a Gaussian (width ~1 keV) and a 5000-K atmosphere with 20% ionization (adapted from Murphy et al. [5]).

The 0.511 MeV emission can originate from different environments at different times during an event. Figure 3 shows the profile of the 0.511 MeV line during the 2003 October 28 eruption, considered to be an extreme event. In this event, the profile was quite wide during the first 2 min of the event compared to the last 16 min. Detailed calculations by Murphy et al. [5] revealed that the early part of the gamma-ray emission (broad line) might have occurred in an environment with a temperature in the range (3–4) ×$10^5$ K and densities ≳ $10^{15}$ cm$^{-3}$. This implies that temperature in the chromosphere has transition-region values. Even though a 6000-K Vernazza et al. [22] model could marginally fit the observations, it was ruled out based on other considerations such as the low atmospheric density inferred (~2 ×$10^{13}$ cm$^{-3}$). On the other hand, the narrow line late in the event is consistent with an environment in which the temperature is very low (~5000 K), the density is same as before (~$10^{15}$ cm$^{-3}$), but the ionization fraction in the gas is ~20%. These results point to the inhomogeneous and dynamic nature of the chromosphere inferred from other considerations [23]. The derived conditions also depend on the atmospheric model, which itself has been revised [24].

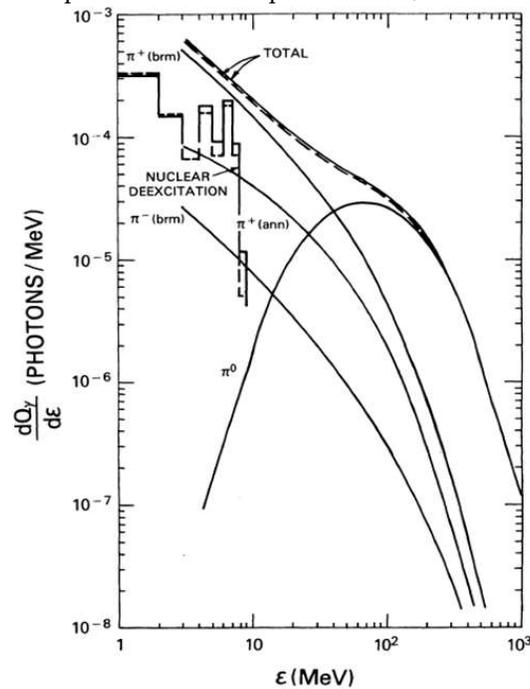

**Figure 4.** Contributions (in units of photons per unit energy interval) to the solar gamma-ray spectrum from $\pi^0$ decay, bremsstrahlung due to electrons from $\pi^-$ decay ($\pi^-$ brm), bremsstrahlung due to positrons from $\pi^+$ decay ($\pi^+$ brm), annihilation radiation of positrons from $\pi^+$ decay ($\pi^+$ ann). The energetic particles responsible for the production of pions were taken to be accelerated from a shock. The top two curves (solid and dashed) represent the total "giraffe" spectrum that combines these four components. The solid curve is for the ambient atmosphere while the dashed curve is for the abundance of the 1982 June 3 event derived from the associated solar energetic particle (SEP) event. The nuclear deexcitation line spectrum for the two abundances are also superposed. (Adapted from Murphy et al. [17]).

*3.2. Pion Continuum*

The pion continuum described briefly above is shown in Figure 4 with different components: (i) the $\pi^0$ decay continuum, which has a characteristic peak around 68 MeV, (ii) the bremsstrahlung continuum due to positrons emitted by $\mu^+$ resulting from $\pi^+$ decay ($\pi^+$ brm), (iii) the positron annihilation continuum due to in-flight annihilation of positrons from $\pi^+$ decay ($\pi^+$ ann), and (iv) the bremsstrahlung continuum due to electrons emitted by $\mu^-$ resulting from $\pi^-$ decay ($\pi^-$ brm). The sum of the four components (Total) represents the spectrum of gamma-rays resulting from pion decay assuming that the accelerated particle angular distribution is isotropic. The $\pi^0$ continuum dominates at energies >100 MeV and determines the spectrum at these energies but its contribution is very tiny at 10 MeV. For example, the $\pi^-$ bremsstrahlung has four times larger contribution than from $\pi^0$ decay, while $\pi^+$ bremsstrahlung and annihilation contributions are larger by factors of 20 and 75, respectively. Below 10 MeV, the gamma-ray spectrum is mostly determined by $\pi^+$ bremsstrahlung. Thus, below ~30 MeV, the combined contribution from positrons dominate the spectrum. The $\pi^0$ continuum exceeds the $\pi^+$ bremsstrahlung around 30 MeV, producing the characteristic "giraffe" shoulder around this energy. Such a spectrum was first derived from the observations of pion continuum in the 1982 June 3 event by Forrest et al. [25], who identified a gamma-ray emission component that lasted for ~20 minutes beyond the impulsive phase of the flare. The spectrum in Figure 4 was calculated in great detail by Murphy et al. [17] to explain the 1982 June 3 event assuming that the primary protons have a shock spectrum [26]. They also performed a similar calculation assuming a stochastic particle spectrum thought to be produced in the reconnection site. While the early part of the 1982 June 3 gamma-rays can be explained by the steep stochastic spectrum, the part extending beyond the impulsive phase (late part) needs to be explained by the shock spectrum, which is much harder than the stochastic spectrum. Furthermore, these authors found that the shock spectrum is similar to the SEP spectrum observed in space.

The time-extended gamma-ray emission first detected by Forrest et al. [25] using the Gamma-Ray Spectrometer on board the Solar Maximum Mission has been observed by many different missions, but such events were rare [27,28]. Two events had durations exceeding ~2 h [29,30]. The Large Area Telescope (LAT) on the Fermi satellite has detected dozens of such time-extended gamma-ray events from the Sun at energies >100 MeV, thanks to the detector's high sensitivity [31]. The average duration of these gamma-ray events is about 9.7 h and with six events lasting for more than 12 h [32,33], including an event that lasted for almost a day. The time-extended events are known by different names "long duration gamma-ray flare (LDGRF)" [34,35], "sustained gamma-ray emission (SGRE)" [33,36], and "late-phase gamma-ray emission (LPGRE)" [32]. The Fermi/LAT observations have revived the interest in the origin of the high-energy particles in these events because the accelerator needs to inject >300 MeV ions toward the chromosphere/photosphere to produce the pions required for the gamma-ray events.

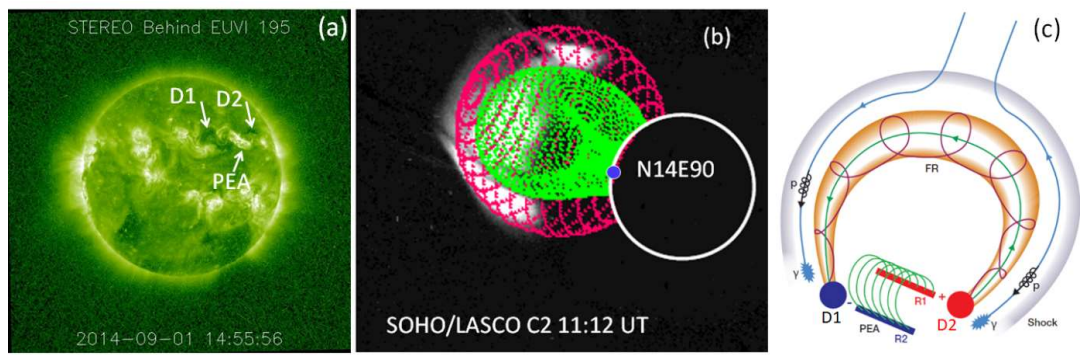

**Figure 5.** (**a**) An extreme ultraviolet image obtained by the STEREO mission showing the spatial structure of the eruption region consisting of dimming regions (D1, D2) and the post-eruption arcade (PEA). (**b**) A flux rope (green) and shock (red) structures superposed on a SOHO white-light image showing the CME. The blue dot (at heliographic coordinates N14E90) represents the centroid of the Fermi/LAT gamma-ray source located between the flux rope (FR) leg and the shock front. The flux rope legs are rooted in the dimming regions D1 and D2. (**c**) A schematic showing the FR and the surrounding shock. Particles accelerated near the shock nose travel along magnetic field lines in the space between the FR and shock, precipitate in the chromosphere/photosphere to produce gamma-rays via the pion-decay mechanisms discussed in the text (adapted from Gopalswamy et al. [40]).

Works focusing on the time-extended nature of these gamma-ray events explore ways to extend the life of the >300 MeV protons from stochastic (impulsive-phase) acceleration, e.g., by particle trapping in flare loops (e.g., [37]). In this scenario, the largest spatial extent of the gamma-ray source is the size of the post-eruption arcade (or flare area) discerned from coronal images taken in extreme ultraviolet wavelengths. In the shock scenario, the gamma-ray source is spatially extended because the angular extent of the shock is much larger than that of the flare structure [38,39]; shock acceleration is naturally time-extended evidenced by type II radio bursts and SEP events [33].

Gopalswamy et al. [40] demonstrated the spatially extended nature of the gamma-ray source during the 2014 September 1 event (see also [36,39,41]). They used multiview data from the Solar and Heliospheric Observatory (SOHO) and the Solar Terrestrial Relations Observatory (STEREO) missions to obtain a detailed picture of the eruption, including magnetic structures that extend beyond the flare structure (post-eruption arcade, PEA) as described in Figure 5. It must be noted that both the flux rope and PEA are products of the eruption process (magnetic reconnection). The PEA remains anchored to the solar surface, while the flux rope is ejected into the heliosphere with speeds exceeding 2000 km/s. The flux rope is a large structure rapidly expanding into the heliosphere compared to the compact flare structure. The flux rope drives a shock because of its high speed and the shock accelerates the required >300 MeV protons. The protons travel down to the chromosphere/photosphere along the field lines located between the flux rope and shock and produce the gamma-rays. Particles traveling away from the shock into the heliosphere are detected as SEP events. Shocks are known to accelerate particles as they propagate into the heliosphere beyond Earth's orbit, but the high-energy particles required for pion production may be accelerated only to certain distance from the Sun; this distance determines the duration of an SGRE event. In the case of the 2014 September 1 event, the SGRE lasted for about 4 hours. With a shock speed >2300 km/s obtained from coronagraph observation, one can infer that the shock stopped accelerating >300 MeV protons to sufficient numbers by the time it reached a distance of about 50 solar radii from the Sun. Evidence for the shock shown in Figure 5 is the interplanetary type II radio burst that lasted until the end of the SGRE event and a bit beyond. The estimated duration of the type II burst (about 7.5 hours) is in agreement with the linear relation found between the two durations [33].

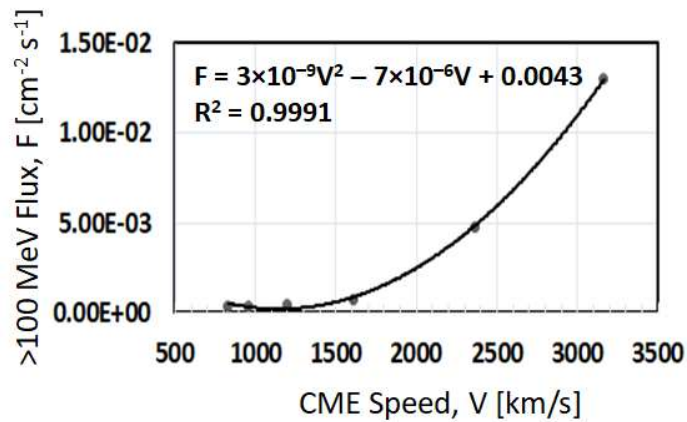

**Figure 6.** Scatter plot between >100 MeV gamma-ray flux (F) from Fermi/LAT against the CME speed (V) for events in which the CME onset was during the impulsive phase. The speeds of three CMEs are different from those in Minasyants et al. [42]. The polynomial fit to the data points and the square of the correlation coefficient (R) are shown on the plot.

In a given eruption, both flare and shock populations are expected to be present, the flare particles being the earliest. Murphy et al. [17] concluded that the nuclear deexcitation line flux is primarily due to the flare population, while the 0.511-line flux has roughly equal contributions from flare and shock populations. On the other hand, the extended phase emission is solely due to the shock population. The conclusion on the extended phase emission initially derived from studying the 1982 June 3 event, seems to be applicable to all events with time-extended emission [33]. Recently, Minasyants et al. [42] found that in certain gamma-ray events with high fluxes of >100 MeV photons, the development of the flare and CME are simultaneous. The CME started during the impulsive phase of the flare. They found that the >100 MeV flux is highly correlated with the CME speed, although the sample is small. Figure 6 shows the relation, replotted with a second-order polynomial fit instead of their linear fit. Normally, one would have thought the impulsive phase gamma-ray flux should be related to the impulsive-phase proton population, and not to the shock population. On further examination, it is found that a type II radio burst started in the impulsive phase of the events, indicating early shock formation. This is typical of eruptions that produce ground level enhancement (GLE) in SEP events [43–45], implying particle acceleration by shocks to GeV energies within the impulsive phase. This result suggests that the shock population may also have contribution to >100 MeV photons in the impulsive phase.

STEREO observations have revealed that shocks can form very close to the Sun, as close as 1.2 solar radii [46]. Shocks typically take several minutes to accelerate particles to GeV energies after their formation. This means, shocks are present within the closed field regions of the corona early on, sending particles toward the Sun and augmenting the impulsive phase particles. The precipitation sites are expected to be different from the PEA as discussed in Figure 5. Once the shock propagates beyond ~2.5 solar radii, accelerated particles can move both ways, away and toward the Sun.

## 4. Conclusions

Positrons are important particles both in the laboratory and in astrophysics. They are extremely useful in understanding high-energy phenomena on the Sun. They provide information on various processes starting from particle acceleration, transport, and interaction with the dense part of the solar atmosphere. Positrons provide information on the physical conditions in the chromosphere/photosphere where they are produced and destroyed by different processes, leaving tell-tale signatures in the gamma-ray spectrum. In addition, gamma-rays and positrons provide information on the magnetic structures involved in solar eruptions that support the acceleration and transport of the highest-energy particles in the inner heliosphere. Spatially resolved gamma-ray observations beyond what is currently available (e.g., [47]) are needed to resolve the issue of the relative contributions from stochastic and shock accelerations in solar eruptive events.

**Funding:** This research was funded by NASA's Living with a Star Program.

**Acknowledgments.** I thank Gerald H. Share and Pertti A. Mäkelä for reading versions of this manuscript and providing useful comments.

**Conflicts of Interest:** The author declares no conflict of interest.

**References**

1. Anderson, C.D. The Positive Electron. *Phys. Rev.* **1933**, *43*, 491–494, doi:10.1103/PhysRev.43.491.
2. Mills, A.P. 2-Positron and Positronium Sources. In *Methods in Experimental Physics*; Dunning, F.B., Hulet, R.G., Eds.; Atomic, Molecular, and Optical Physics: Charged Particles; Academic Press: Cambridge, MA, USA, 1995; Volume 29, pp. 39–68, doi:10.1016/S0076-695X(08)60653-5.
3. Guessoum, N.; Jean, P.; Gillard, W. The Lives and Deaths of Positrons in the Interstellar Medium. *Astron. Astrophys.* **2005**, *436*, 171–185, doi:10.1051/0004-6361:20042454.
4. Knödlseder, J.; Jean, P.; Lonjou, V.; Weidenspointner, G.; Guessoum, N.; Gillard, W.; Skinner, G.; von Ballmoos, P.; Vedrenne, G.; Roques, J.-P.; et al. The All-Sky Distribution of 511 KeV Electron-Positron Annihilation Emission. *Astron. Astrophys.* **2005**, *441*, 513–532, doi:10.1051/0004-6361:20042063.
5. Murphy, R.J.; Share, G.H.; Skibo, J.G.; Kozlovsky, B. The Physics of Positron Annihilation in the Solar Atmosphere. *Astrophys. J. Suppl. Ser.* **2005**, *161*, 495–519, doi:10.1086/452634.
6. Gopalswamy, N. Properties of Interplanetary Coronal Mass Ejections. *Space Sci. Rev.* **2006**, *124*, 145–168, doi:10.1007/s11214-006-9102-1.
7. Aschwanden, M.J. Particle Acceleration and Kinematics in Solar Flares—A Synthesis of Recent Observations and Theoretical Concepts (Invited Review). *Space Sci. Rev.* **2002**, *101*, 1–227, doi:10.1023/A:1019712124366.
8. Wild, J.P.; Smerd, S.F.; Weiss, A.A. Solar Bursts. *Annu. Rev. Astron. Astrophys.* **1963**, *1*, 291, doi:10.1146/annurev.aa.01.090163.001451.
9. Forbush, S.E. Three Unusual Cosmic-Ray Increases Possibly Due to Charged Particles from the Sun. *Phys. Rev.* **1946**, *70*, 771–772, doi:10.1103/PhysRev.70.771.
10. Morrison, P. Solar Origin of Cosmic Ray Time Variations. *Il Nuovo Cimento* **1958**, *7*, 858–865.

11. Elliot, H. The Nature of Solar Flares. *Planet. Space Sci.* **1964**, *12*, 657–660, doi:10.1016/0032-0633(64)90024-8.
12. Lingenfelter, R.E.; Ramaty, R. High Energy Nuclear Reactions in Solar Flares. *High-Energy Nucl. React. Astrophys.* **1967**, *99*.
13. Chupp, E.L.; Forrest, D.J.; Higbie, P.R.; Suri, A.N.; Tsai, C.; Dunphy, P.P. Solar Gamma Ray Lines Observed during the Solar Activity of 2 August to 11 August 1972. *Nature* **1973**, *241*, 333–335, doi:10.1038/241333a0.
14. Kozlovsky, B.; Lingenfelter, R.E.; Ramaty, R. Positrons from Accelerated Particle Interactions. *Astrophys. J.* **1987**, *316*, 801–818, doi:10.1086/165244.
15. Kozlovsky, B.; Murphy, R.J.; Share, G.H. Positron-Emitter Production in Solar Flares from 3He Reactions. *Astrophys. J.* **2004**, *604*, 892–899, doi:10.1086/381969.
16. Ramaty, R.; Murphy, R.J. Nuclear Processes and Accelerated Particles in Solar Flares. *Space Sci. Rev.* **1987**, *45*, 213–268, doi:10.1007/BF00171995.
17. Murphy, R.J.; Dermer, C.D.; Ramaty, R. High-Energy Processes in Solar Flares. *Astrophys. J. Suppl. Ser.* **1987**, *63*, 28, doi:10.1086/191180
18. Benz, A.O. Flare Observations. *Living Rev. Sol. Phys.* **2016**, *14*, 2, doi:10.1007/s41116-016-0004-3.
19. Ramaty, R.; Mandzhavidze, N. Solar Flares: Gamma Rays. *Encyclopedia of Astronomy and Astrophysics, Edited by Paul Murdin, article id. 2292. Bristol: Institute of Physics Publishing,* **2001**, *1–4*
20. Share, G.H.; Murphy, R.J.; Smith, D.M.; Schwartz, R.A.; Lin, R.P. RHESSI e+-e- Annihilation Radiation Observations: Implications for Conditions in the Flaring Solar Chromosphere. *Astrophys. J. Lett.* **2004**, *615*, L169–L172, doi:10.1086/426478.
21. Ley, R. Atomic Physics of Positronium with Intense Slow Positron Beams. *Appl. Surf. Sci.* **2002**, *194*, 301–306, doi:10.1016/S0169-4332(02)00139-3.
22. Vernazza, J.E.; Avrett, E.H.; Loeser, R. Structure of the Solar Chromosphere. III - Models of the EUV Brightness Components of the Quiet-Sun. *Astrophys. J. Suppl. Ser.* **1981**, *45*, 635–725, doi:10.1086/190731.
23. Song, P. A Model of the Solar Chromosphere: Structure and Internal Circulation. *Astrophys. J.* **2017**, *846*, 92, doi:10.3847/1538-4357/aa85e1.
24. Avrett, E.H.; Loeser, R. Models of the Solar Chromosphere and Transition Region from SUMER and HRTS Observations: Formation of the Extreme-Ultraviolet Spectrum of Hydrogen, Carbon, and Oxygen. *Astrophys. J. Suppl. Ser.* **2008**, *175*, 229–276, doi:10.1086/523671.
25. Forrest, D.J.; Vestrand, W.T.; Chupp, E.L.; Rieger, E.; Cooper, J.F.; Share, G.H. Neutral Pion Production in Solar Flares. *ICRC* **1985**, *146–149*.
26. Ellison, D.C.; Ramaty, R. Shock Acceleration of Electrons and Ions in Solar Flares. *Astrophys. J.* **1985**, *298*, 400–408, doi:10.1086/163623.
27. Chupp, E.L.; Ryan, J.M. High Energy Neutron and Pion-Decay Gamma-Ray Emissions from Solar Flares. *Res. Astron. Astrophys.* **2009**, *9*, 11–40, doi:10.1088/1674-4527/9/1/003.
28. Kuznetsov, S.N.; Kurt, V.G.; Yushkov, B. Yu.; Myagkova, I.N.; Galkin, V.I.; Kudela, K. Protons Acceleration in Solar Flares: The Results of the Analysis of Gamma-Emission and Neutrons Recorded by the SONG Instrument Onboard the CORONAS-F Satellite. *Astrophys. Space Sci. Lib.* **2014**, *400*, 301, doi:10.1007/978-3-642-39268-9_10.
29. Akimov, V.V.; Afanassyey, V.G.; Belaousov, A.S.; Blokhintsev, I.D.; Kalinkin, L.F.; Leikov, N.G.; Nesterov, V.E.; Volsenskaya, V.A.; Galper, A.M.; Chesnokov, V.J.; et al. Observation of High Energy Gamma-Rays from the Sun with the GAMMA-1 Telescope (E > 30 MeV). *ICRC* **1991**, *3*, 73.
30. Kanbach, G.; Bertsch, D.L.; Fichtel, C.E.; Hartman, R.C.; Hunter, S.D.; Kniffen, D.A.; Kwok, P.W.; Lin, Y.C.; Mattox, J.R.; Mayer-Hasselwander, H.A. Detection of a Long-Duration Solar Gamma-Ray Flare on 11 June 1991 with EGRET on COMPTON-GRO. *Astron. Astrophys. Suppl. Ser.* **1993**, *97*, 349–353.
31. Atwood, W.B.; Abdo, A.A.; Ackermann, M.; Althouse, W.; Anderson, B.; Axelsson, M.; Baldini, L.; Ballet, J.; Band, D.L.; Barbiellini, G.; et al. The Large Area Telescope on the Fermi Gamma-Ray Space Telescope Mission. *Astrophys. J.* **2009**, *697*, 1071–1102, doi:10.1088/0004-637X/697/2/1071.
32. Share, G.H.; Murphy, R.J.; White, S.M.; Tolbert, A.K.; Dennis, B.R.; Schwartz, R.A.; Smart, D.F.; Shea, M.A. Characteristics of Late-Phase $\greater$100 MeV Gamma-Ray Emission in Solar Eruptive Events. *Astrophys. J.* **2018**, *869*, 182, doi:10.3847/1538-4357/aaebf7.
33. Gopalswamy, N.; Mäkelä, P.; Yashiro, S.; Lara, A.; Akiyama, S.; Xie, H. On the Shock Source of Sustained Gamma-Ray Emission from the Sun. *J. Phys. Conf. Ser.* **2019**, *1332*, 012004, doi:10.1088/1742-6596/1332/1/012004.
34. Ryan, J.M. Long-Duration Solar Gamma-Ray Flares. *Space Sci. Rev.* **2000**, *93*, 581–610.

35. de Nolfo, G.A.; Bruno, A.; Ryan, J.M.; Dalla, S.; Giacalone, J.; Richardson, I.G.; Christian, E.R.; Stochaj, S.J.; Bazilevskaya, G.A.; Boezio, M.; et al. Comparing Long-Duration Gamma-Ray Flares and High-Energy Solar Energetic Particles. *Astrophys. J.* **2019**, *879*, 90, doi:10.3847/1538-4357/ab258f.
36. Plotnikov, I.; Rouillard, A.P.; Share, G.H. The Magnetic Connectivity of Coronal Shocks from Behind-the-Limb Flares to the Visible Solar Surface during γ-Ray Events. *Astron. Astrophys.* **2017**, *608*, A43, doi:10.1051/0004-6361/201730804.
37. Ryan, J.M.; Lee, M.A. On the Transport and Acceleration of Solar Flare Particles in a Coronal Loop. *Astrophys. J.* **1991**, *368*, 316, doi:10.1086/169695.
38. Cliver, E.W.; Kahler, S.W.; Vestrand, W.T. On the Origin of Gamma-Ray Emission from the Behind-the-Limb Flare on 29 September 1989. *ICRC* **1993**, *3*, 91.
39. Pesce-Rollins, M.; Omodei, N.; Petrosian, V.; Liu, W.; Rubio da Costa, F.; Allafort, A.; Fermi-LAT Collaboration. Fermi Large Area Telescope Observations of High-Energy Gamma-Ray Emission from behind-the-Limb Solar Flares. *ICRC* **2015**, *34*, 128.
40. 40. Gopalswamy, N.; Mäkelä, P.; Yashiro, S.; Akiyama, S.; Xie, H.; Thakur, N. Source of Energetic Protons in the 2014 September 1 Sustained Gamma-Ray Emission Event. *Sol. Phys.* **2020**, *295*, 18, doi:10.1007/s11207-020-1590-8.
41. Jin, M.; Petrosian, V.; Liu, W.; Nitta, N.V.; Omodei, N.; Rubio da Costa, F.; Effenberger, F.; Li, G.; Pesce-Rollins, M.; Allafort, A.; et al. Probing the Puzzle of Behind-the-Limb γ-Ray Flares: Data-Driven Simulations of Magnetic Connectivity and CME-Driven Shock Evolution. *Astrophys. J.* **2018**, *867*, 122, doi:10.3847/1538-4357/aae1fd.
42. Minasyants, G.; Minasyants, T.; Tomozov, V. Features of Development of Sustained Fluxes of High-Energy Gamma-Ray Emission at Different Stages of Solar Flares. *Sol.-Terr. Phys.* **2019**, *5*, 10–17, doi:10.12737/stp-53201902.
43. Reames, D.V. Solar Energetic-Particle Release Times in Historic Ground-Level Events. *Astrophys. J.* **2009**, *706*, 844–850, doi:10.1088/0004-637X/706/1/844.
44. Gopalswamy, N.; Xie, H.; Yashiro, S.; Akiyama, S.; Mäkelä, P.; Usoskin, I.G. Properties of Ground Level Enhancement Events and the Associated Solar Eruptions during Solar Cycle 23. *Space Sci. Rev.* **2012**, *171*, 23–60.
45. Gopalswamy, N.; Xie, H.; Akiyama, S.; Yashiro, S.; Usoskin, I.G.; Davila, J.M. The First Ground Level Enhancement Event of Solar Cycle 24: Direct Observation of Shock Formation and Particle Release Heights. *Astrophys. J. Lett.* **2013**, *765*, L30, doi:10.1088/2041-8205/765/2/L30.
46. Gopalswamy, N.; Xie, H.; Mäkelä, P.; Yashiro, S.; Akiyama, S.; Uddin, W.; Srivastava, A.K.; Joshi, N.C.; Chandra, R.; Manoharan, P.K.; et al. Height of Shock Formation in the Solar Corona Inferred from Observations of Type II Radio Bursts and Coronal Mass Ejections. *Adv. Space Res.* **2013**, *51*, 1981–1989, doi:10.1016/j.asr.2013.01.006.
47. Omodei, N.; Pesce-Rollins, M.; Longo, F.; Allafort, A.; Krucker, S. Fermi-LAT Observations of the 2017 September 10 Solar Flare. *Astrophys. J. Lett.* **2018**, *865*, L7, doi:10.3847/2041-8213/aae077.